\documentclass{agujournal}

\journalname{JGR-Planets}

\begin{document}

%
%

\title{The Liquidus Temperature for Methanol-Water Mixtures at High Pressure
and Low Temperature, with Application to Titan}

%
%




\authors{A. J. Dougherty\affil{1},
	 Z. T. Bartholet\affil{1},
	 R. J. Chumsky\affil{1},
	 K. C. Delano\affil{1}\thanks{Current address, Dept.\ of Physics,
	       U.\ Texas at San Antonio, San Antonio, TX, USA},
	 X. Huang\affil{1},
         D. K. Morris\affil{1}\thanks{Current address, 
	      Division of Geological and Planetary Sciences,
	      California Institute of Technology, Pasadena, CA, USA}
	 }


\affiliation{1}{Department of Physics, Lafayette College, Easton, Pennsylvania 18042 USA}




\correspondingauthor{Andrew Dougherty}{doughera@lafayette.edu}




\begin{keypoints}
\item Titan's icy shell and subsurface ocean may be affected by any impurities such as methanol
\item We measure the freezing point of methanol/water solutions as a function of pressure
\item Previous estimates may have underestimated the effectiveness of methanol as an antifreeze
\end{keypoints}

%
%


\begin{abstract}
Methanol is a potentially important impurity in subsurface oceans on
Titan and Enceladus.
We report measurements of the freezing of methanol-water samples at
pressures up to 350~MPa using a volumetric cell with sapphire windows.
For low concentrations of methanol, the liquidus temperature is typically
a few degrees below the corresponding ice freezing point, while at high
concentrations it follows the pure methanol trend.  In the Ice-III regime,
we observe several long-lived metastable states.  The results suggest
that methanol is a more effective antifreeze than previously estimated,
and might have played an important role in the development of Titan's
subsurface ocean.
\end{abstract}

%
%

\section{Introduction}
%

A number of icy satellites likely contain subsurface oceans \citep{Nimmo2016}.  
On
Titan, evidence for a subsurface ocean comes from several sources,
including gravity measurements \citep{Iess2012} and electrical
measurements of a Schumann resonance \citep{Beghin2012}, as well
as computations based on obliquity \citep{Baland2011} and surface
topography \citep{Hemingway2013, Nimmo2010}.

Detailed models of the interior of Titan rely on the properties of
aqueous solutions at the low temperatures and high pressures found within
such an ocean, and such properties would be affected by the presence of
impurities.  Important impurities likely include ammonia, but methanol
may also be present \citep{Kargel1992,Hogenboom1997} and is included in models
such as \citet{Neveu2017}, \citet{Dunaeva2016}, \citet{Lefevre2014}, \citet{Fortes2012},
\citet{Tobie2012}, \citet{Davies2010}, and \citet{Deschamps2010}.  More broadly,
methanol is also an important volatile found in the outer solar system nebula \citep{Sekine2014},
on the surfaces of some trans-Neptunian objects \citep{Merlin2012}, and 
in some star-forming regions \citep{Dartois1999,Shimonishi2016}.

As a potent antifreeze compound, small concentrations of methanol
could play an important role in the formation and development
of the subsurface ocean \citep{Deschamps2010}.  In addition,
the rheology of methanol-containing slurries could have significant
implications for the structure of possible cryovolcanic flows
\citep{Zhong2009,Davies2010,Lopes2007}.
Although there is not conclusive evidence for widespread cryovolcanic
activity on Titan \citep{Moore2011,Lopes2013}, there are regions for which
it may be a reasonable explanation of topographic features
\citep{Lopes2013,Solomonidou2014}, and it is one possible source of
replenishment for Titan's atmospheric methane \citep{Davies2016}.

On Enceladus, the observations of water-rich plumes
\citep{Dougherty2006,Porco2006,Waite2009,Hansen2011,Hsu2015}
reveal either the presence of a liquid reservoir or highly active
melting.  The composition of the plumes suggests a liquid reservoir
\citep{Hsu2015,Postberg2011}, and the measured libration is consistent
with a global subsurface ocean, rather than a localized reservoir
\citep{Thomas2016}.

As with Titan, any subsurface ocean would likely contain impurities,
such as ammonia and methanol \citep{Kargel1992}, that act as powerful
antifreeze compounds.  Small amounts of methanol may have been detected
on the surface of Enceladus \citep{Hodyss2009}, as well as in the
plume \citep{Waite2009}.  In addition to being a powerful antifreeze,
methanol could also play a role in the formation of methane hydrates
\citep{McLaurin2014}.

More generally, thermodynamic models of the interiors of icy moons can
benefit from an improved understanding of how the addition of small
amounts of methanol will change the temperature and pressure-dependent
properties of ice \citep{Deschamps2010, Hsieh2015}.
In the model of \citet{Deschamps2010}, for
example, the authors estimated the pressure-dependent behavior of a
methanol-water solution
by interpolating between the pure water and pure methanol behavior.

In this work, we present results for the liquidus temperature for methanol
concentrations between 5 wt.\% and 75 wt.\%, for pressures from 5 MPa to 350 MPa.
We find that interpolation underestimates the effect
of small concentrations of methanol on the freezing point.


\section{Materials and Methods}

\subsection{Materials}

\begin{figure}[hbt!]
\centering
\noindent\includegraphics[width=0.8 \columnwidth]{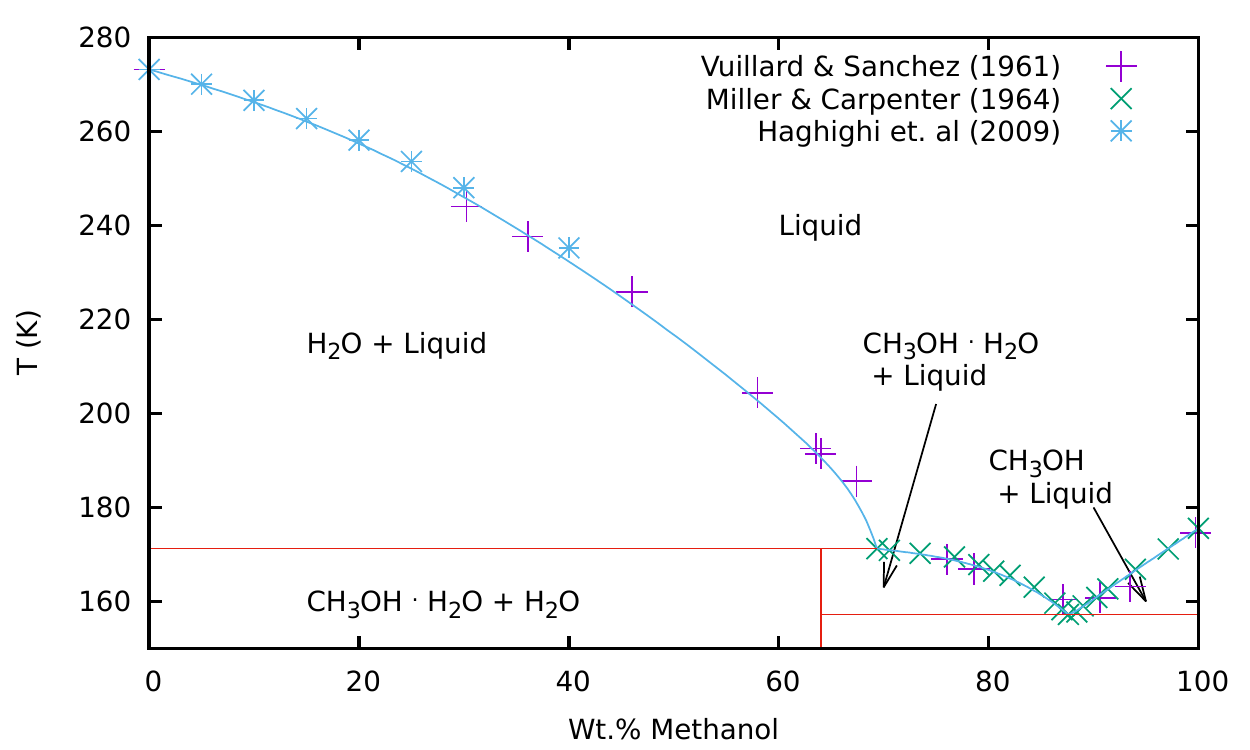}
    \caption{Phase diagram for methanol/water, based on compilations by
    \citet{Haghighi2009}, \citet{Vuillard1961}, and \citet{Miller1964},
    and adapted from \citet{Kargel1992,Ott1979,Takaizumi1997}.
    Phases are solid unless indicated otherwise.
    The hydrated solid phase is methanol monohydrate
    \citep{Fortes2006,Fortes2011}.
    }
    \label{phasediagram}
\end{figure}

The phase diagram for methanol/water solutions at atmospheric pressure
is shown in Fig.~\ref{phasediagram}.  The eutectic concentration is
approximately 88~wt.\%.

Sample solutions were made from HPLC-grade methanol
(EMD Chemicals, 99.8\%) and HPLC-grade filtered water (Fisher Scientific).
Solutions of approximately 4.9, 9.9, 33.8, and 75 wt.\% methanol
were studied.
(All concentrations are methanol mass divided by total solution mass.)
Since methanol is highly hygroscopic, the concentration would change
somewhat during the extensive filling and leak-testing process.  The
beaker of solution was exposed to the atmosphere for periods of hours, and
residual water from flushing and cleaning the connecting tubing could not
be completely eliminated.  In addition, when leak-testing with partially
frozen Ice Ih samples, any leaked fluid would be enriched in methanol.
The net effect could be a significant change, on the order of 4\%.
Accordingly, in each case, once the system was filled and sealed,
and leak testing was completed, the concentration of the sample was
estimated by extrapolating the low-pressure liquidus temperature to 1
atmosphere, and comparing it with the data in Fig.~\ref{phasediagram}.
Those estimated concentrations are the values reported here.

The range of concentrations was chosen to explore both
the low-concentration regime where the initial solid formed is ice Ih,
and the high-concentration regime where the initial solid formed is
methanol monohydrate.  The former is relevant for the proposed
primordial bulk composition of Titan's ocean \citep{Deschamps2010}, while
the latter would be relevant for the freezing of a trapped subsurface
body of liquid, where the concentration would tend towards the eutectic.

The range of pressures was limited by the maximum pressure able to be
maintained with the apparatus, generally about 350~MPa.  For models
of Titan that include an outer Ice Ih shell roughly 100~km thick and
a subsurface ocean atop an inner high-pressure Ice V or Ice VI layer,
such as \citet{Fortes2012} and \citet{Dunaeva2014}, this would correspond
to the pressure in the ocean at a depth of roughly 120~km below
the Ice Ih shell.

\subsection{Methods}

\begin{figure}[hbt!]
\centering
\noindent\includegraphics[width=0.9 \columnwidth]{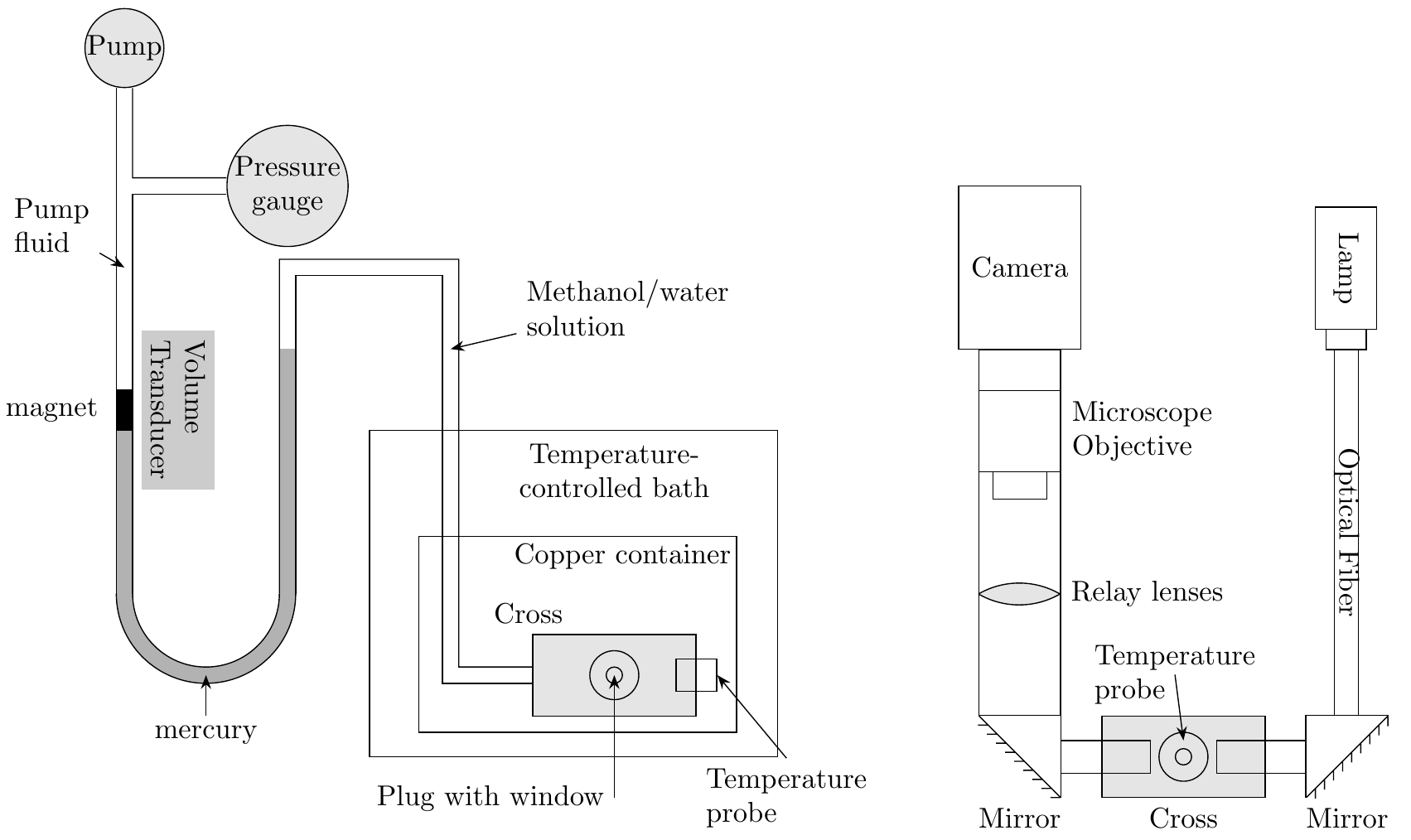}
    \caption{Diagrams of the pressure system (left) and imaging system (right).
    The imaged portion of the sample is confined to a 1~mm-wide gap between
    the two windows at the center of the cross.
    This system allows simultaneous measurements of pressure, temperature,
    and volume changes, along with optical images of the sample.}
    \label{fig:apparatus}
\end{figure}

The apparatus and technique are similar to those described in
previous papers for magnesium sulfate and ammonia-water mixtures
\citep{Hogenboom1995,Hogenboom1997}, enhanced to allow optical access
to the sample.  The system is shown in Fig.~\ref{fig:apparatus}.

Approximately 1 mL of sample is loaded into a pressure cell that is placed
in a copper container and immersed in an insulated, temperature-controlled
ethanol/water bath.  The pressure cell is made from a 316 stainless
steel block with four ports, known as a cross (High Pressure Equipment
Company \#60-HF6).  Two opposing ports contain replaceable plugs that
have sapphire windows sealed in them with epoxy.  The third port contains
a plug in which a silicon diode thermometer is installed.  The fourth
port connects the cell to the pressure system.
Pressure is applied through a pump with ethylene glycol as the pump fluid.
Pressure is measured with two Heise gauges: an analog pressure gauge
valid for all pressures studied, and a more sensitive digital pressure
gauge for pressures below 200~MPa.

The sample in the pressure cell is separated from the ethylene glycol pump
fluid by a vertical U-tube filled with mercury.  A steel capillary tube
of constant cross section forms one arm of the U-tube.  A small Alnico
magnet is placed in the capillary on the interface between the pump fluid
and the mercury, and the height of that magnet is measured by a transducer.
Changes in the transducer voltage are approximately proportional to
changes in sample volume.  As long as the sample is mostly liquid,
this system allows simultaneous measurements of temperature, pressure,
and volume of the sample.

The imaging system consists of a lamp that shines light through an
infrared filter and optical fiber that directs the beam horizontally
through the sample cell.  The infrared filter is used to minimize heating
of the sample by the light source.  After passing through the cell, the
beam is reflected by a $45^\circ$ mirror upward through a matched pair
of lenses to a long working distance optical microscope objective coupled
to a Pulnix digital camera.  The camera obtains images of a vertical
cross-section of the sample, with $ 1392 \times 1040$
pixels and an overall resolution of about 1.7 $\mu$m/pixel.  The gap
between the sapphire windows is approximately 1~mm.
Although the camera's field of view does not cover the entire system,
we typically observed dissolving or growing crystals corresponding to
changes in temperature, pressure, and volume, indicating that the crystal
images reflect the phase transitions within the sample.

For runs above a temperature of 210~K, the temperature is controlled
by a immersion cooler in the ethanol/water bath.  The cooling and
warming rates are controlled by computer.  In these viscous solutions,
it is important to allow sufficient time for the system to approach
thermodynamic equilibrium.  Typically, the temperature was increased
at a rate of 0.005~K/minute, or even more slowly, to ensure that the
system was near thermodynamic equilibrium.  For colder temperatures
(such as were needed for the 75 wt.\% runs),
liquid nitrogen was used to cool the system, and the system warmed up as
the nitrogen evaporated.  The warming rates in those cases were typically
closer to 0.1~K/minute.


\section{Data}

\begin{figure}[hbt!]
\centering
\noindent\includegraphics[width=0.8 \columnwidth]{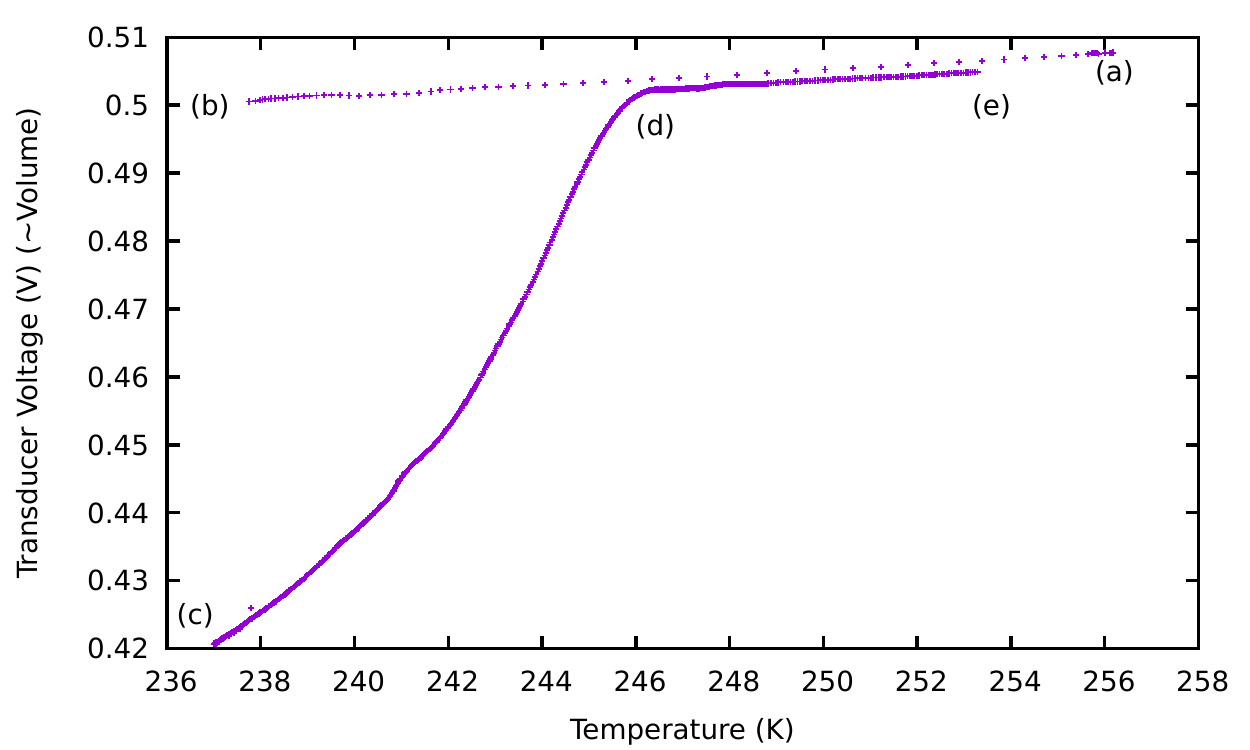}
\caption{Data for a run with 4.9 wt.\% methanol at a nominal pressure of
263~MPa.  The vertical axis shows transducer voltage (which varies approximately
linearly with volume), while the horizontal axis shows temperature.  The labelled
points are described in the text.  The liquidus temperature for this run
was 245.9~K.}
\label{fig:265MPa}
\end{figure}

The results for a run with a 4.9 wt.\% solution at a nominal pressure of
263~MPa are shown in Fig.~\ref{fig:265MPa}.  The horizontal axis shows
temperature, while the vertical axis gives the transducer voltage,
which varies approximately linearly with volume.  The run started with
a homogeneous liquid (point a) at a temperature of 256~K.  Upon cooling,
the liquid contracted gradually until about 237.8~K (point b), where the
supercooled sample rapidly froze and many ice crystals were visible in
the window.  This crystallization was accompanied by a sharp decrease in
volume and pressure, indicating that the solid formed was denser than
the surrounding liquid.  Upon further cooling, the pressure reached
about 252~MPa at a temperature of 237~K (point c).  The sample was then
gradually warmed at a rate of 0.005~K/min.  As the sample warmed, the
volume gradually increased and ice crystals could be observed falling
downward in the image.  At 245.9~K (point d), the melting was complete,
and the now-liquid sample gradually expanded with temperature as it
continued to warm (point e).  A typical run would last about three days.

The liquidus temperature was then estimated by finding the intersection of
straight lines fitted to the end of the melting curve (just before point
d) and the pure liquid warming curve (after point d).  For sufficiently
slow warming, the transition was fairly sharp and the uncertainties in the
liquidus temperature were typically less than $\pm 0.2$~K.  For transition
temperatures below about 210~K, where liquid nitrogen was used for
cooling and the warming rates were closer to 0.1~K/min, the transitions
were more rounded and the uncertainties were closer to $\pm 1.0$~K.
Uncertainties in pressure were typically less than $\pm 0.3$~MPa for
runs below 200~MPa, and about $\pm 1$~MPa for runs above 200~MPa, due
to the different pressure gauges used.

\begin{figure}[hbt!]
\centering
\noindent\includegraphics[width=0.5 \columnwidth]{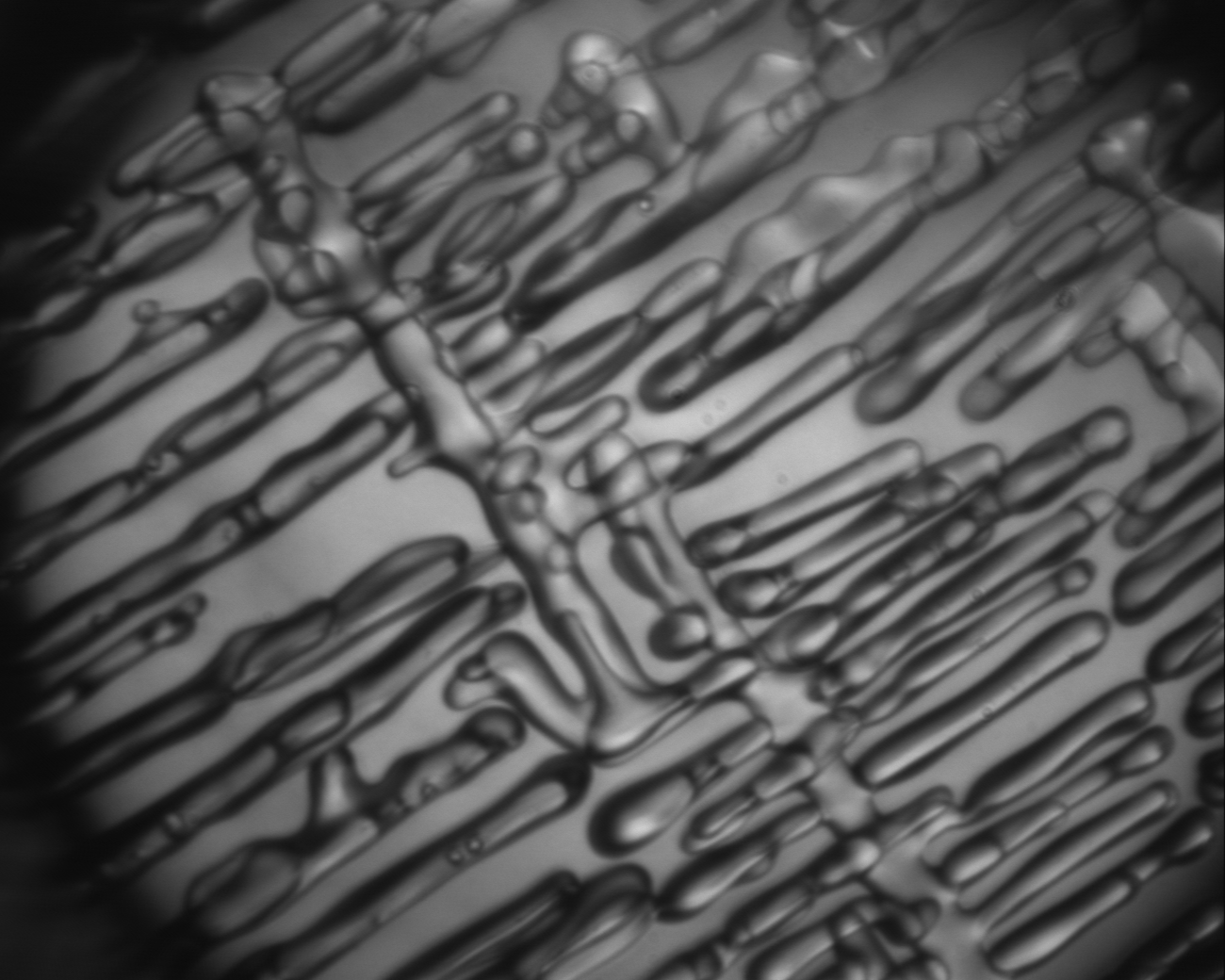}
\caption{Ice-III crystals in a 4.9 wt.\% methanol solution slowly
    melting at a pressure of 212.8~MPa and temperature of
    245.5~K (about 2~K below the liquidus temperature).  The image
    is approximately 2.2~mm across.
    Gravity points downward.  As the
    crystal melts, the arms detach and fall downward, indicating that
    the solid is more dense than the surrounding liquid.}
    \label{fig:dendrites}
\end{figure}

An image of crystals formed at 212.8~MPa, presumably Ice-III, is shown in
Fig.~\ref{fig:dendrites}.  The volume decreased when these crystals
formed, indicating that the solid is denser than the surrounding liquid.
Upon warming, the volume increased and the crystals fell downward, again
indicating a denser solid phase.

\section{Results}

\subsection{Low methanol concentrations}

\begin{figure}[hbt!]
\centering
\noindent\includegraphics[width=0.8 \columnwidth]{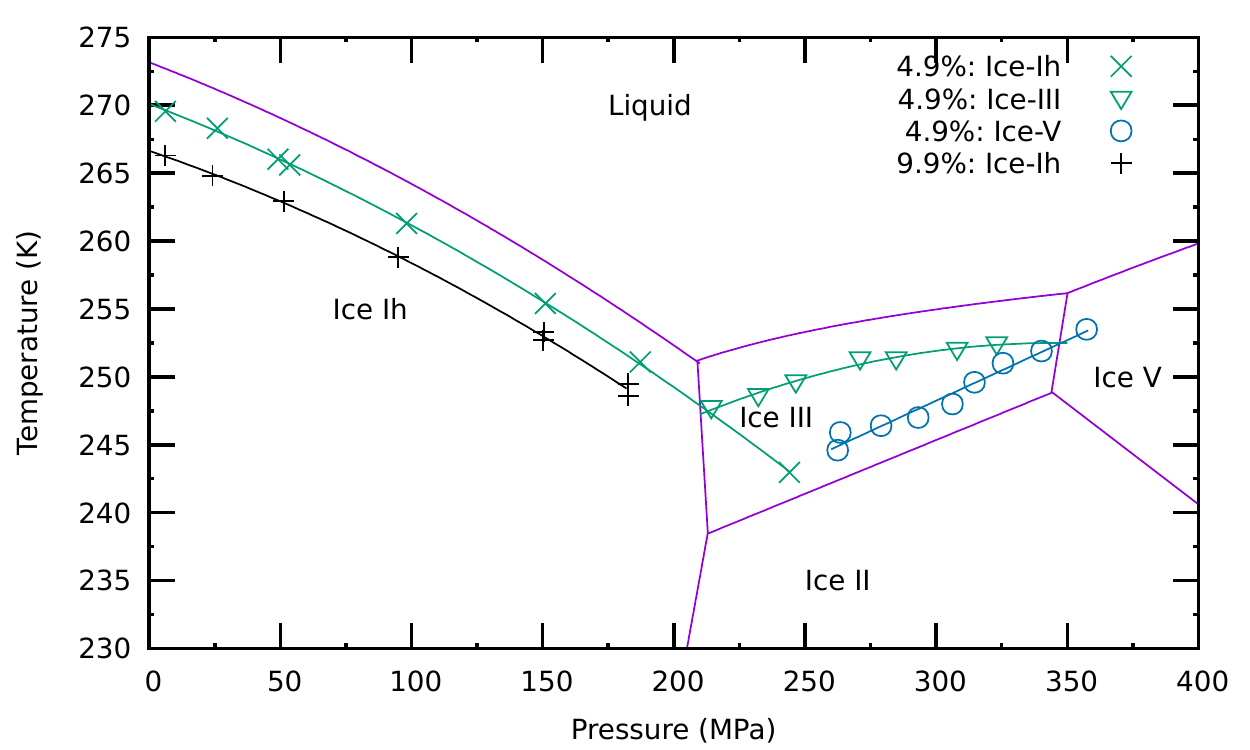}
\caption{Liquidus temperatures for two low-concentration samples as
a function of pressure.  Shown for comparison are the pure ice melting
curves from \citet{Wagner2011} and the phase boundaries for the different
ice phases from \citet{Dunaeva2010}.  The uncertainties in the data
points are typically smaller than the symbols on the graph.  The solid
lines through the data are fits described in the text.}
\label{fig:methanol-low}
\end{figure}

The resulting liquidus temperatures for two low-concentration samples
(4.9~wt.\% and 9.9~wt.\%) are shown in Fig.~\ref{fig:methanol-low},
along with the boundaries for various phases of ice.  (For the 9.9\% sample,
a leak in the pressure system prevented the acquisition of data at pressures
above about 200~MPa.)

In the region between 200~MPa and 350~MPa, we observed three different
states, all for the same 4.9~wt.\% sample.  In one run at 244~MPa, the
volume increased upon freezing, and the ice crystals were less dense
than the surrounding fluid.  The liquidus point for that sample was
approximately 243~K, clearly consistent with a metastable continuation
of the Ice-Ih regime curve.
We also observed two different states where the volume decreased upon
freezing, and the ice crystals were more dense than the surrounding fluid.
These states fall on two distinct curves in Fig.~\ref{fig:methanol-low}.
The states labeled Ice-III had a lower density than the ones labeled Ice-V,
but we were not able to measure absolute densities of solids with this
apparatus.  The lower-temperature data may follow an extension of the Ice
V regime curve, consistent with the formation of metastable Ice V, as
suggested by \citet{Evans1967}, and are included in the Ice V regime fits.

The two states also differed in the experimental protocol necessary for
their formation.  For pressures below
about 240~MPa, the system would spontaneously freeze upon supercooling
into either Ice-Ih or Ice-III (as in Fig.~\ref{fig:dendrites}).
For pressures above 240~MPa, the system would spontaneously freeze as
Ice-V (as in Fig.~\ref{fig:265MPa}).  In order to prepare Ice-III samples
at higher pressures, we would first cool at a pressure of approximately
230~MPa until ice crystals formed, and then raise the pressure up to
the desired target.

\subsection{Fits}
To fit the data, we considered a number of different candidate functional forms.
In particular, we tried
functions of the form
used by \citet{Wagner2011} for modeling the pure ice melting curves,
\begin{linenomath*}
\begin{equation}
\centering
\frac{p}{p_0} = 1 +
\sum_1^3 a_i \left(1 - \left(\frac{T}{T_0}\right)^{b_i}\right)
\label{eqn:pfit}
\end{equation}
\end{linenomath*}
where the numerical values of the parameters for pure ice are given
in \citet{Wagner2011}.

However, our data do not cover a wide enough range of temperatures to
constrain the parameters, particularly in the Ice III regime, especially
since the the values for $p_0$ and $T_0$ are not independently known.

We also considered the Simon-Glatzel equation \citep{Simon1929},
as well as the modified form proposed by \citet{Kechin1995}:
\begin{linenomath*}
\begin{equation}
\frac{T}{T_0} = \left(1 + \frac{\Delta p}{a_1} \right)^{a_2}
    \exp\left(- a_3 \Delta p \right)
\label{eqn:kechin}
\end{equation}
\end{linenomath*}
where $\Delta p = p - p_0$.
This is similar to including a single term from Eq.~\ref{eqn:pfit},
but with an additional exponential decay term $a_3$.  However, that
term did not significantly improve the quality of the fit, and the
values for $a_3$ were consistent with zero.  We also considered other
forms sometimes used for the melting of pure substances, such as that
recommended by \citet{Yi-Jing1982}, but they either did not fit as well
over the full range of the data, or required more parameters that were
typically poorly constrained.

The best descriptions of the data were ultimately obtained with simple
polynomial fits of the form
\begin{linenomath*}
\begin{equation}
T = a_0 + a_1 (p - p_0) + a_2 (p - p_0)^2
\label{eqn:poly}
\end{equation}
\end{linenomath*}
The $p_0$ term was
arbitrarily fixed at a convenient pressure so that the
uncertainties more accurately reflected the fit over the
pressure range of interest.  The coefficients for the curves in
Fig.~\ref{fig:methanol-low} are given in Table~\ref{tbl:methanol-low}.


\begin{table}[htb!]
\caption{Coefficients of fits to Eq.~\ref{eqn:poly} for various curves in
Fig.~\ref{fig:methanol-low}, along with the corresponding temperature ranges
of the data.}
\label{tbl:methanol-low}
\centering
\begin{tabular}{c l l r c c c}
\hline
Conc. & Regime & T Range (K) & $p_0$ (MPa) & $a_0$ (K) & $a_1$ (K/MPa) & $a_2$ (K/MPa$^2$) \\
\hline
4.9\%  & Ice Ih & 248--270 & 0 &
    $ 270.1  \pm 0.1 $ &
    $ -0.074 \pm  0.003 $ &
    $ -(1.5 \pm 0.1) \times 10^{-4} $  \\
4.9\%  & Ice III & 248--252 & 209 &
    $ 247.2 \pm  0.3 $ &
    $ 0.078 \pm  0.011 $ &
    $ -(2.9 \pm 0.9) \times 10^{-4} $  \\
4.9\%  & Ice V & 245--254 & 350 &
    $ 252.7 \pm  0.3 $ &
    $ 0.089 \pm  0.006 $ &
          --               \\
9.9\%  & Ice Ih & 249--266.5 & 0 &
    $ 266.6 \pm  0.3 $ &
    $ -0.066  \pm 0.009 $ &
    $ -(1.65 \pm 0.5) \times 10^{-4} $  \\
\hline
\end{tabular}
\end{table}

\subsection{Higher methanol concentrations}

\begin{figure}[hbt!]
\centering
\noindent\includegraphics[width=0.8 \columnwidth]{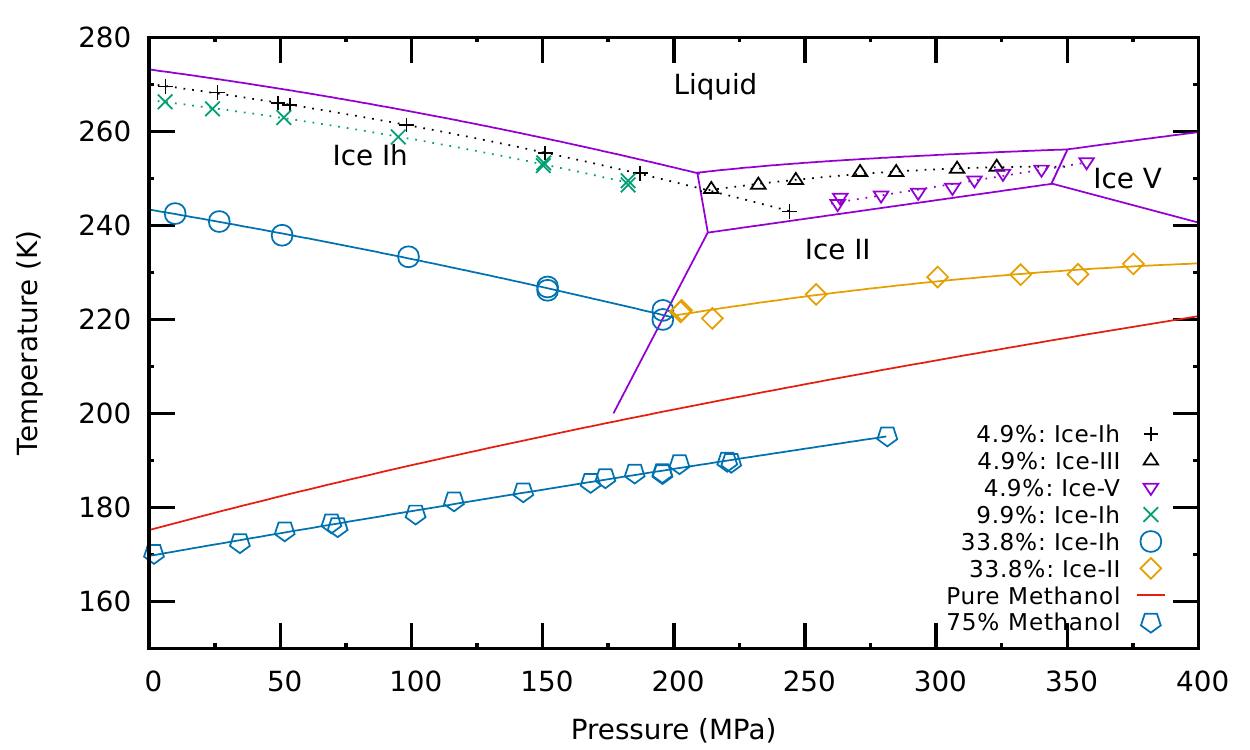}
\caption{Liquidus temperatures for four different concentrations
as a function of pressure.  The fits are described in the text.
In addition to the data shown in
Fig.~\ref{fig:methanol-low}, the melting curve of pure methanol
\citep{Sun1988, Wurflinger1977,Gromnitskaya2004} is included for
comparison.}
\label{fig:methanol-full}
\end{figure}

The results for two additional samples, at
33.8 and 75 wt.\%, are included in Fig.~\ref{fig:methanol-full}.
For the 75 wt.\% sample, the solid formed is presumably
CH$_3$OH$\cdot$H$_2$O, as shown in the phase diagram
(Fig.~\ref{phasediagram}).
The coefficients for the fits
are given in Table~\ref{tbl:methanol-full}.

\begin{table}[htb!]
\caption{Coefficients of fits to Eq.~\ref{eqn:poly} for the high-concentration
curves in Fig.~\ref{fig:methanol-full}}
\label{tbl:methanol-full}
\centering
\begin{tabular}{c l r c c c}
\hline
Concentration & Regime & $p_0$ (MPa) & $a_0$ (K) & $a_1$ (K/MPa) & $a_2$ (K/MPa$^2$) \\
\hline
33.8\%  & Ice Ih & 0 &
    $ 243.4 \pm 0.6 $ &
    $ -0.095 \pm 0.016 $ &
    $ -(9.7 \pm 7.5) \times 10^{-5} $  \\
33.8\%  & Ice II & 209 &
    $ 221.6 \pm 0.6 $ &
    $ 0.087 \pm 0.023 $ &
    $ -(1.7 \pm 1.5) \times 10^{-4} $  \\
75\%  &  & 0 &
    $ 169.7 \pm 0.4 $ &
    $ 0.098 \pm 0.006 $ &
    $ -(2.7 \pm 2.3) \times 10^{-5} $  \\
\hline
\end{tabular}
\end{table}

%

\section{Discussion}

Methanol is a potentially important antifreeze.  Specifically, we
find that the freezing point depression due to the addition of methanol
is about 3.1~K for a concentration of 4.9~wt.\%, 5.8~K for 9.9~wt.\%,
and grows to about 31~K for a concentration of 33.8~wt.\%.
For the low
concentrations of planetary relevance ($< 10$~wt.\%), we find the freezing point
depression $fz(x)$ (in Kelvin) due to the addition of methanol is roughly
independent of pressure, at least
for the two concentrations measured (4.9~wt.\% and 9.9~wt.\%)
in the Ice-Ih regime, and
is reasonably represented over that range as a function of percent methanol
concentration $x$ by
\begin{linenomath*}
\begin{equation}
fz(x) = (0.59 \pm 0.01) x  .
\label{eqn:fzdp}
\end{equation}
\end{linenomath*}

In contrast, a linear interpolation between the Ice-Ih and pure methanol
curves, as was used in \citet{Deschamps2010}, gives a freezing point depression
that decreases with increasing pressure, and hence tends to underestimate the
effect of adding methanol.  Specifically, for a 4.9 wt.\% concentration of
methanol and a pressure of 200~MPa, the measured liquidus temperature based
on Table~\ref{tbl:methanol-low} yields a freezing point depression of
approximately 3.1~K, while a linear interpolation yields a smaller effect of
only 1.6~K.  Or, equivalently, that same 1.6~K freezing point depression
at 200~MPa could be achieved with only a 2.7~wt.\% concentration.

\section{Conclusions}

In their modeling of Titan's primordial ocean, \citet{Deschamps2010}
found that the inclusion of methanol reduced the solidification
temperature, and hence reduced convection and heat transfer in the outer
Ice Ih shell.  Ultimately, this could lead to an end to crystallization
and the maintenance of a subsurface ocean.  The main contribution from
the present experiments is to observe that methanol is an even more
effective antifreeze than expected, so that a subsurface ocean might be
maintained with an even lower methanol concentration than 
considered there.

More generally, models of the interiors of Titan, Enceladus, and other bodies
with subsurface oceans can benefit from more information about the
thermodynamic behaviors of aqueous solutions at the relevant
temperatures and pressures.  For Titan and Enceladus, methanol is one
reasonable impurity to include in such models.
For other outer solar system bodies and even exoplanets, incorporation of 
methanol may similarly lead to a subsurface ocean being maintained 
where a pure water ocean might completely freeze.
Since many of the of thermodynamic
and rheological properties of an ocean and icy shell are strongly 
temperature-dependent, 
it may be worthwhile to include this freezing point depression
in future modeling.

\clearpage

%
%
%
%
%
%
%

\acknowledgments
Financial support was provided by Lafayette College.  Data used in
the figures is available at
https://data.mendeley.com/datasets/fwpf6t3bxn/1.
The authors thank D.L.\ Hogenboom for
assistance in building the pressure system.

\listofchanges

\end{document}